\documentclass[12pt]{article}
\usepackage{psfig}
\usepackage{axodraw}
\usepackage{amsfonts}
\begin{document}

\def\lesssim{\mathrel{\mathpalette\vereq<}}
\def\gtrsim{\mathrel{\mathpalette\vereq>}}
\makeatletter
\def\vereq#1#2{\lower3pt\vbox{\baselineskip1.5pt \lineskip1.5pt
\ialign{$\m@th#1\hfill##\hfil$\crcr#2\crcr\sim\crcr}}}
\makeatother

\newcommand{\rem}[1]{{\bf #1}}
\newcommand{\tev}{~{\rm TeV}}
\newcommand{\gev}{~{\rm GeV}}
\newcommand{\mev}{~{\rm MeV}}
\newcommand{\kev}{~{\rm keV}}
\newcommand{\ev}{~{\rm eV}}
\newcommand{\cm}{~{\rm cm}}
\newcommand{\vev}[1]{\langle #1 \rangle}
\renewcommand{\thefootnote}{\fnsymbol{footnote}}
\setcounter{footnote}{0}
\begin{titlepage}
\begin{center}

%
\hfill
UW-PT-01/09\\
\hfill
UCB-PTH-01/12\\
\hfill
LBNL-47677\\
\vskip .5in                                                                    
{\Large \bf GUT Breaking on the Brane\footnote{
This work was supported in part by the U.S. Department of 
Energy under Contracts DE-AC03-76SF00098, in part 
by the National Science Foundation under grant PHY-95-14797.}}
\vskip .40in                                                                   
Yasunori Nomura$^{1}$\footnotemark\footnotetext{Research fellow,
  Miller Institute for Basic Research in Science.}, 
David Smith$^{1}$ and 
Neal Weiner$^{2}$
\vskip .20in
$^{1}${\em Department of Physics,}\\
and \\
 {\em Theoretical Physics Group, 
  Lawrence Berkeley National Laboratory,\\
  University of California, Berkeley, California 94720}
\vskip 0.08in
$^{2}${\em Department of Physics,\\
  University of Washington, Seattle, Washington 98195, USA}
\end{center}
\vskip .5in
\begin{abstract}
We present a five-dimensional supersymmetric SU(5) theory in which
the gauge symmetry is broken maximally 
({\it i.e.} at the 5D Planck scale $M_*$) 
on the same 4D brane where chiral matter is localized.  Masses of
the lightest Kaluza-Klein modes for the colored Higgs and $X$ and $Y$
gauge fields are determined by the compactification scale of the fifth
dimension, $M_C \sim 10^{15}$ GeV, rather than by $M_*$.  These fields'
wave functions are repelled from the GUT-breaking brane, so that
proton decay rates are suppressed below experimental limits.  Above
the compactification scale, the {\it differences} between the
standard model gauge couplings evolve logarithmically, so that
ordinary logarithmic gauge coupling unification is preserved.
The maximal breaking of the grand unified group can also lead to other 
effects, such as $O(1)$ deviations from SU(5) predictions of Yukawa
couplings, even in models utilizing the Froggatt-Nielsen mechanism.
\end{abstract}                                                                 

\end{titlepage}                                                                
\renewcommand{\thepage}{\arabic{page}}
\setcounter{page}{1}
\renewcommand{\thefootnote}{\arabic{footnote}}
\setcounter{footnote}{0}
\setcounter{footnote}{0}
\setcounter{table}{0}
\setcounter{figure}{0}

\section{Introduction}

The standard model of particle physics, described by the gauge group 
SU(3)$\otimes$SU(2)$\otimes$U(1) is one of the most successful 
physical theories ever. Nonetheless, the standard model is 
theoretically unsatisfying for a number of reasons. For instance, 
the instability 
of the weak scale against radiative corrections has motivated the 
study of numerous theories, including technicolor, supersymmetry and 
theories with additional dimensions.

While the gauge hierarchy may be the most compelling motivation for new 
physics, there are additional reasons to consider 
theories with further structure. It was long ago realized that the 
gauge group of the standard model could be embedded into a simple 
group \cite{Georgi:1974sy}. The appeal of this idea was substantiated by 
measurements of electroweak observables which suggested that 
the values of the standard model gauge couplings unify at a 
scale $M_{\rm GUT}\sim 10^{15}\gev$ \cite{Georgi:1974yf}.

As data became increasingly precise, it became apparent that 
the simplest grand unified theory (GUT), minimal SU(5), 
predicted a value for $\sin^2\theta_W$ that was
incompatible with observation.   However, the combination of 
supersymmetry and grand unification \cite{Dimopoulos:1981zb} has 
met with great success in predicting $\sin^2\theta_W$ 
\cite{Dimopoulos:1981yj}.  For these supersymmetric theories, 
the GUT symmetry must be broken at a somewhat higher 
scale, $M_{\rm GUT}\sim 2 \times 10^{16}\gev$. 

One smoking gun signal of grand unification is proton 
decay, which arises from dimension six operators 
generated by exchange of the $X$ and $Y$ gauge 
bosons, and, in supersymmetric theories, from dimension five operators 
generated by colored Higgsino exchange \cite{Sakai:1982pk}. 
In supersymmetric GUTs, 
the dimension six operators typically do not lead to observable proton 
decay, but the dimension five operators do, and
place strong limits on these theories \cite{Hisano:1993jj}.
Recently Ref.~\cite{aaronhitoshigutlimits} studied the dimension five 
proton decay and found that even large masses for the first-two 
generation sparticles cannot save minimal supersymmetric SU(5). 
They found that successful gauge coupling unification 
requires that the Higgs triplet mass fall in the range 
$1.1 \times 10^{14}\gev \le M_{H_{C}} \le 9.3 \times 10^{15}\gev$. 
However, even with decoupling of the superpartners, 
the limit from proton decay was shown to require 
$M_{H_{C}} \ge 9.4 \times 10^{16}\gev$.

While additional field content may generate threshold effects to rectify 
this, in this paper we propose an alternative possibility. We will embed a 
unified gauge theory into five dimensions in which gauge and Higgs 
fields propagate in the additional dimension. We will show that if the 
GUT is broken on a brane at a high scale ({\it i.e.}, $M_*$, the 
Planck scale of the five-dimensional theory), the wave functions of 
the gauge fields and Higgs triplets can develop an approximate node 
on the brane, suppressing proton decay operators arising from their 
exchange. We will show that the running above the compactification 
scale is not only consistent with, but is in fact equivalent to 
ordinary four dimensional logarithmic unification with 
the compactification scale identified as the mass of the Higgs triplet.

In doing so, we reformulate the question of the $M_{\rm GUT}/M_{\rm Pl}$ 
hierarchy because the breaking occurs at the fundamental scale of the 
theory (which, in general, is still larger than $10^{16}\gev$). 
In its place we require a 
somewhat large hierarchy $(O(100))$ between the Planck and 
compactification scales. The traditional GUT scale, $2\times 10^{16}\gev$, 
is a derived scale in this framework: there is no new physics there.
Because of the $O(1)$ breaking of the grand unified group on the 
brane, we will also easily understand deviations 
from SU(5) predicted relationships among Yukawa couplings.

\section{Framework}

We consider a 5D supersymmetric SU(5) theory with the fifth 
dimension compactified on an $S_1/Z_2$ orbifold 
(in section \ref{sec:SO(10)} we will consider the possibility 
 of larger gauge groups).  The bulk exhibits N=1 
supersymmetry in five dimensions, which translates into an N=2 
supersymmetry in four dimensions.  The fields that propagate in the
bulk are contained in a gauge supermultiplet and two Higgs 
hypermultiplets transforming as ${\bf 5}$ and ${\overline {\bf 5}}$ 
under SU(5).  Under the 4D N=1 supersymmetry preserved after the
orbifold compactification, the gauge supermultiplet decomposes into
a vector superfield $V$ and an associated 
chiral adjoint $\Phi$. The two hypermultiplets yield four chiral 
multiplets $H_{\bf 5}$, $H_{\bf 5}^{c}$, $H_{\overline{\bf 5}}$ and 
$H_{\overline{\bf 5}}^{c}$. ($H_{\bf 5}$ and $H_{\overline{\bf 5}}^{c}$ 
transform as ${\bf 5}$ under the SU(5), while $H_{\bf 5}^{c}$ and 
$H_{\overline{\bf 5}}$ as $\overline{\bf 5}$.)  
Under the orbifold $Z_2$, which flips the sign of the fifth coordinate 
($y \rightarrow -y$), the various superfields transform as
\begin{eqnarray}
  H_{{\bf 5},\overline{\bf 5}}
    \longrightarrow H_{{\bf 5},\overline{\bf 5}} &\hskip 0.5in &
  H_{{\bf 5},\overline{\bf 5}}^{c}
    \longrightarrow -H_{{\bf 5},\overline{\bf 5}}^{c} \nonumber\\
  V \longrightarrow V &\hskip 0.5in& \Phi \longrightarrow -\Phi,
\end{eqnarray}
which leaves zero modes only for the MSSM fields and their GUT 
counterparts.

The $S_1/Z_2$ orbifold has fixed points at $y=0$ and $y=\pi R$.
We take chiral matter to be localized to the $y=\pi R$ fixed
point, along with an adjoint chiral superfield $\Sigma$.  
The SU(5) symmetry is broken by the vacuum
expectation value (vev) of $\Sigma$, which we assume to have the form
\begin{equation}
        {\vev {\bf \Sigma_{ij}}} = \pmatrix{
        2/5 & 0 & 0 & 0 & 0 \cr
        0 & 2/5 & 0 & 0 & 0 \cr
        0 & 0 & 2/5 & 0 & 0 \cr
        0 & 0 & 0 & -3/5 & 0 \cr
        0 & 0 & 0 & 0 & -3/5} \frac{\vev{\Sigma}}{\sqrt{2}}.
\end{equation}
We imagine that this breaking occurs at or near the five-dimensional 
Planck scale $M_*$, which we assume to be much larger 
than the conventional GUT scale of $2\times 10^{16}\gev$.
We will refer to the $y=\pi R$ fixed point as the GUT-breaking brane.

Bulk fields even under the orbifold $Z_2$ can couple directly to 
the fields on the GUT-breaking brane.  For instance, the Higgs fields 
can couple to the chiral matter fields $T_{\bf 10}$ and 
$F_{\overline{\bf 5}}$ as $(\lambda_u T_{\bf 10} T_{\bf 10} H_{\bf 5} 
+ \lambda_d T_{\bf 10} F_{\overline{\bf 5}} H_{\overline{\bf 5}})
\delta(y-\pi R)$ in the superpotential.  We also take 
there to be a bare Higgs mass term, $\alpha H_{\bf 5} H_{\overline{\bf 5}} 
\delta(y-\pi R)$, as well as Higgs coupling to the GUT breaking, 
$(\beta/M_{*}) H_{\bf 5} \Sigma H_{\overline{\bf 5}} \delta(y-\pi R)$, 
in the superpotential.  While it would be interesting to study standard 
solutions to the doublet-triplet splitting problem in 
this framework, we will not endeavor to do so here. 
Rather, we will simply tune these two contributions against 
one another so that the SU(2) doublets $H_u$ $(\subset H_{\bf 5})$ and 
$H_d$ $(\subset H_{\overline{\bf 5}})$ are nearly massless, 
while the colored triplets $H_C$ $(\subset H_{\bf 5})$ and 
$H_{\overline C}$ $(\subset H_{\overline{\bf 5}})$ end up with a 
large mass term $\kappa {H}_C {H}_{\overline C} \delta(y-\pi R)$. 
Note that $\kappa$ is a dimensionless parameter since $H_C$ and 
$H_{\overline C}$ are bulk fields and have dimension $3/2$.
Here, $\kappa$ arises as a sum of terms (with higher order terms 
potentially relevant in the strongly coupled theory). In addition to 
the bare mass term, the $H_{\bf 5} \Sigma H_{\bf \overline 5}$ term 
contributes. The natural size of this operator 
is restricted only by perturbativity, with an upper
bound $\sim 6 \pi^2$ suggested by naive dimensional analysis 
in higher dimensions \cite{Chacko:2000hg}.

\subsection{Spectrum: Gauge fields}
Once the gauge symmetry is broken, the $X$ and $Y$ bosons acquire
brane-localized masses through the interaction
\begin{equation}
  \mathcal{L} \supset \delta(y-\pi R) g_{5}^{2} \vev{\Sigma}^{2} 
    A_{\mu,\hat{a}}A^{\mu}_{\hat{a}}
\end{equation}
where $\hat{a}$ indexes the broken generators of the group and $g_5^2$ is
the 5D gauge coupling, with mass dimension $-1$. If we were to  
decompose each $X$ and $Y$ bulk vector field in a naive Kaluza-Klein (KK) 
basis, we would estimate that the zero modes have mass $M_{V}=g_5
\vev{\Sigma}/\sqrt{2 \pi R} =g_{4} \vev{\Sigma}$. 
Such an estimate incorrectly assumes that the gauge boson wave 
functions do not change appreciably in the 
presence of the large localized mass term. 
To find the correct spectrum we must 
solve the differential equation
\begin{equation}
  -\partial_y^{2} A_{\mu} + \delta(y-\pi R) g_{5}^{2} \vev{\Sigma}^{2}
    A_{\mu} = m^2 A_{\mu},
\end{equation}
which leads to KK masses ${M_n^G}$ given by 
\cite{Chacko:2000hg,Arkani-Hamed:2001mi}
\begin{equation}
  {M_n^G}  \tan({M_n^G} \pi R) = \frac{g_{5}^{2} \vev{\Sigma}^{2}}{2},
\label{eq:gaugespectrum}
\end{equation}
where $n = 0,1,2,\cdots$.
For $g_{5}^{2} \vev{\Sigma}^{2} R \gg 1$ this equation gives 
a spectrum whose low-lying levels are approximately
$M_n^G = (n+1/2) M_{C}$ (with $M_C \equiv 1/R$), whereas 
the usual KK spectrum is $m_{n}= n M_{C}$. 
We see that the KK tower has been shifted up one half unit.

This is easy to understand intuitively. The gauge field picks up a 
mass of $g_{4}\vev{\Sigma} \gg 1/R$ if it does not avoid the brane. If 
it avoids the brane entirely, it does not ``see'' the mass term, and 
picks up a smaller mass of $1/(2R) = M_{C}/2$ from a nontrivial profile 
of the wave function in the extra dimension. The localized mass term 
acts merely to dynamically assign the boundary condition that the 
wave functions should vanish at the GUT-breaking brane. 
The boundary condition is not absolute, of course, so that
the spectrum is not precisely spaced according to $(n+1/2)M_{C}$. 
The value of the wave function at the GUT-breaking brane goes as 
$\cos({M_n^G}\pi R) = (2n+1)M_C/(g_{5}^{2}\vev{\Sigma}^{2})$, and
there are corrections to the mass which go like 
$- (2n+1)M_C^2/(\pi g_{5}^{2}\vev{\Sigma}^{2})$.
Modes with masses near or above the scale $g_{5}^2 \vev{\Sigma}^2$ become
essentially degenerate with the ordinary KK tower, $m_n = n M_C$. 

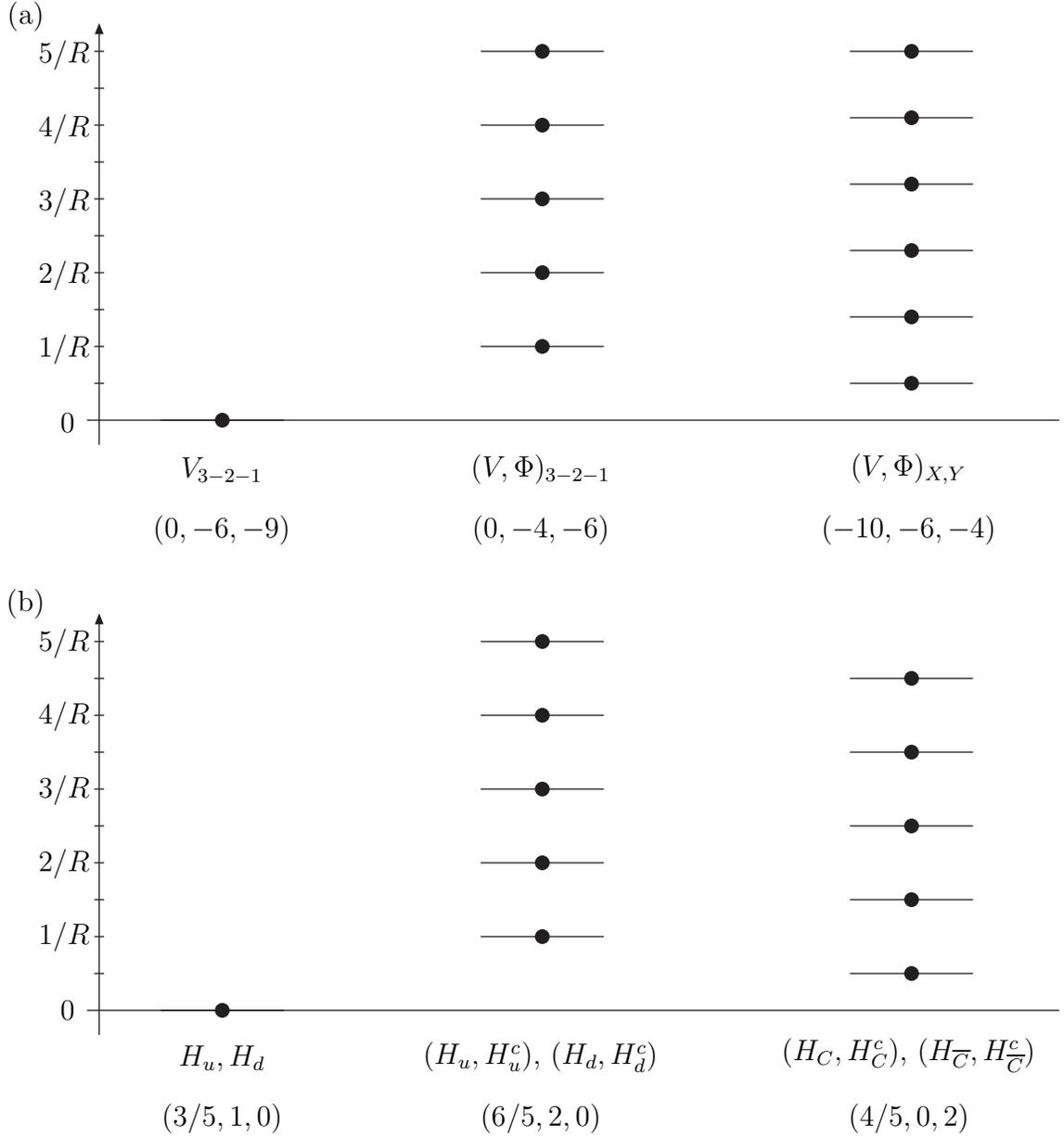
\begin{figure}
\begin{center} 
\begin{picture}(400,460)(-20,-60)
  \Text(-20,400)[b]{(a)}  
  \Line(5,240)(400,240)
  \LongArrow(10,230)(10,400)
  \Text(0,240)[r]{$0$}
  \Line(8,255)(12,255)    
  \Line(8,270)(12,270)    \Text(6,270)[r]{$1/R$}
  \Line(8,285)(12,285)    
  \Line(8,300)(12,300)    \Text(6,300)[r]{$2/R$}
  \Line(8,315)(12,315)  
  \Line(8,330)(12,330)  \Text(6,330)[r]{$3/R$}
  \Line(8,345)(12,345)    
  \Line(8,360)(12,360)    \Text(6,360)[r]{$4/R$}
  \Line(8,375)(12,375)  
  \Line(8,390)(12,390)  \Text(6,390)[r]{$5/R$}
  \Text(60,215)[b]{$V_{3-2-1}$}
  \Text(60,190)[b]{$(0,-6,-9)$}
  \Line(35,240)(85,240)      \Vertex(60,240){3}
  \Text(190,215)[b]{$(V,\Phi)_{3-2-1}$}\Text(190,190)[b]{$(0,-4,-6)$}
  \Line(165,270)(215,270)    \Vertex(190,270){3}
  \Line(165,300)(215,300)    \Vertex(190,300){3}
  \Line(165,330)(215,330)  \Vertex(190,330){3}
  \Line(165,360)(215,360)    \Vertex(190,360){3}
  \Line(165,390)(215,390)  \Vertex(190,390){3}
  \Text(340,215)[b]{$(V,\Phi)_{X,Y}$}\Text(340,190)[b]{$(-10,-6,-4)$} 
  \Line(315,255)(365,255)    \Vertex(340,255){3}
  \Line(315,282)(365,282)    \Vertex(340,282){3}
  \Line(315,309)(365,309)  \Vertex(340,309){3}
  \Line(315,336)(365,336)    \Vertex(340,336){3}
  \Line(315,363)(365,363)  \Vertex(340,363){3}
  \Line(315,390)(365,390)  \Vertex(340,390){3}
  \Text(-20,160)[b]{(b)}
  \Line(5,0)(400,0)
  \LongArrow(10,-10)(10,160)
  \Text(0,0)[r]{$0$}
  \Line(8,15)(12,15)    
  \Line(8,30)(12,30)    \Text(6,30)[r]{$1/R$}
  \Line(8,45)(12,45)    
  \Line(8,60)(12,60)    \Text(6,60)[r]{$2/R$}
  \Line(8,75)(12,75)  
  \Line(8,90)(12,90)  \Text(6,90)[r]{$3/R$}
  \Line(8,105)(12,105)    
  \Line(8,120)(12,120)    \Text(6,120)[r]{$4/R$}
  \Line(8,135)(12,135)  
  \Line(8,150)(12,150)  \Text(6,150)[r]{$5/R$}
  \Text(60,-25)[b]{$H_u, H_d$}
  \Text(60,-50)[b]{$(3/5,1,0)$}
  \Line(35,0)(85,0)      \Vertex(60,0){3}
  \Text(190,-25)[b]{$(H_u, H_u^c)$, $(H_d, H_d^c)$}
  \Text(190,-50)[b]{$(6/5,2,0)$}
  \Line(165,30)(215,30)    \Vertex(190,30){3}
  \Line(165,60)(215,60)    \Vertex(190,60){3}
  \Line(165,90)(215,90)  \Vertex(190,90){3}
  \Line(165,120)(215,120)    \Vertex(190,120){3}
  \Line(165,150)(215,150)  \Vertex(190,150){3}
  \Text(340,-25)[b]{$(H_C, H_C^c)$, $(H_{\overline C}, H_{\overline C}^c)$}
  \Text(340,-50)[b]{$(4/5,0,2)$}
  \Line(315,15)(365,15)    \Vertex(340,15){3}
  \Line(315,45)(365,45)    \Vertex(340,45){3}
  \Line(315,75)(365,75)  \Vertex(340,75){3}
  \Line(315,105)(365,105)    \Vertex(340,105){3}
  \Line(315,135)(365,135)  \Vertex(340,135){3}
\end{picture}
\caption{
Mass spectrum for the lowest KK modes of the gauge fields (a), and
Higgs fields (b) in our model.  As explained in the text, the transition 
to degeneracy between the $X,Y$ and 3-2-1 towers occurs much more 
gradually than shown in (a).  In (b), the limit of very large 
$\kappa$ is taken, so that the slight non-degeneracy between the 
colored hypermultiplet pairs at each level is not resolved. In both 
(a) and (b), the triplet of numbers below each tower corresponds to 
the beta function contribution $(b_1,b_2,b_3)$ that
comes from each level in that tower (for the colored Higgs, the
contributions from the nearly-degenerate hypermultiplet pairs are
combined).}
\label{fig:spectrum}
\end{center}
\end{figure}

We have not broken supersymmetry, and thus the gauginos corresponding
to the broken SU(5) generators have an 
identical spectrum.  The KK towers for the gauge fields are
illustrated in Fig.~\ref{fig:spectrum}a.  For each unbroken
SU(3)$\otimes$SU(2)$\otimes$U(1) (3-2-1) generator
there is a massless vector multiplet and massive vector multiplets at
$M_C$, $2M_C$, $3M_C$, and so on.  For each broken ($X,Y$) generator
there is a tower of massive vector multiplets whose low levels are
offset from those of the 3-2-1 tower by $M_C/2$, but whose higher
levels relax into alignment with the 3-2-1 tower.  Although in
Fig.~\ref{fig:spectrum}a this transition to degeneracy has already
been achieved at $5M_C$, for the parameters of interest in our model,
the relaxation will actually occur much more gradually, over the span
of $\sim g_5^2 \vev{\Sigma}^2/M_C \sim O(10-100)$ KK levels.  Note
that for each $X,Y$ massive vector multiplet but one there is a
corresponding  3-2-1 massive vector multiplet.  The remaining one
is paired with the {\em massless} 3-2-1 vector multiplet, indicating
a larger total number of states in the $X,Y$ tower relative to the
3-2-1 tower.  These excess states are the eaten Goldstone degrees of
freedom from the $\Sigma$ multiplet.  This means that above the scale 
$\sim g_5^2 \vev{\Sigma}^2$ the KK towers of gauge fields contributing 
to the renormalization group (RG) running are completely SU(5) symmetric 
except that one chiral adjoint for each 3-2-1 vector multiplet is missing.  
These missing degrees of freedom are provided by the physical 
$\Sigma$ fields which remain after the Goldstone 
components of $\Sigma$ are eaten by the $X,Y$ gauge multiplets.
Therefore, above the scale $g_5^2 \vev{\Sigma}^2$ and $m_\Sigma$ 
(the mass of physical $\Sigma$ fields), the spectrum contributing to 
the running is completely SU(5) symmetric.

\subsection{Spectrum: Higgs fields}
The spectrum  analysis for the Higgs triplet fields $H_C$ and
$H_{\overline C}$ is somewhat different, because 
the brane coupling $\kappa$ is dimensionless, and hence the naive mass
that the zero modes would pick up by not avoiding the brane is
just $\kappa M_{C}$ which is comparable to $M_{C}$ for order one $\kappa$. 
For our purposes, we will take $\kappa$ to be a large parameter, roughly
$O(10)$, so that the Higgs wave functions will in fact be strongly
repelled from the brane. This size for $\kappa$
is quite consistent with the general features of the theory, which must 
be somewhat strongly coupled in order to achieve order one top 
Yukawa and gauge couplings in spite of the large radius.

The equations that determine the colored Higgsino spectrum are
\begin{equation}
  -i \overline \sigma^{\mu}\partial_{\mu} H_{C} - \partial_{y} \overline 
  H_{C}^{c} - \kappa \delta(y-\pi R) \overline H_{{\overline C}}=0,
\label{eq:higgsinoeqs1}
\end{equation}
\begin{equation}
  -i \overline \sigma^{\mu}\partial_{\mu} H_{\overline C} - \partial_{y} 
  \overline H_{\overline C}^{c} - \kappa \delta(y-\pi R) 
  \overline H_{C}=0, \nonumber
\label{eq:higgsinoeqs2}
\end{equation}
\begin{equation}
  -i \overline \sigma^{\mu}\partial_{\mu} H_{C}^{c}+ \partial_{y} \overline 
  H_{C}=0, \nonumber
\label{eq:higgsinoeqs3}
\end{equation}
\begin{equation}
  -i \overline \sigma^{\mu}\partial_{\mu} H_{\overline C}^{c} 
  + \partial_{y} \overline H_{\overline C}=0. \nonumber
\label{eq:higgsinoeqs4}
\end{equation}
We can analyze these equations by writing the solutions as
\begin{eqnarray}
  H_{C}^{(c)}&=& \sum_n \eta_{C,n}^{(c)}(x) g_{C,n}^{(c)}(y), \\
  H_{\overline C}^{(c)}&=& 
    \sum_n \eta_{{\overline C},n}^{(c)}(x) g_{{\overline C},n}^{(c)}(y),
\end{eqnarray}
where $g_{C,n}$ and $g_{{\overline C},n}$ are even under 
$y \rightarrow -y$, while $g_{C,n}^c$ and $g_{{\overline C},n}^c$ are odd.
If we integrate the first two of these equations in a region $(\pi 
R - \epsilon, \pi R + \epsilon)$ where $\epsilon \rightarrow 0$, 
we obtain two constraints,
\begin{equation}
  \eta_{C,n}^{c}(x)=\frac{\kappa g_{{\overline C},n}(\pi R)}{2
    g_{C,n}^{c}(\pi R - \epsilon)} \eta_{{\overline C},n}(x),
\end{equation}
\begin{equation}
  \eta_{{\overline C},n}^{c}(x)=\frac{\kappa g_{C,n}(\pi R)}{2
    g_{{\overline C},n}^{c}(\pi R - \epsilon)}\eta_{C,n}(x).
\end{equation}
Here, two constraints must be satisfied mode by mode.
Let us define
\begin{equation}
  \gamma_{(I,J),n}=\frac{\kappa g_{I,n}(\pi R)}
    {2 g_{J,n}^{c}(\pi R - \epsilon)},
\end{equation}
where $I,J$ take $C, {\overline C}$, and identify both 
$-i \overline \sigma^{\mu}\partial_{\mu} \eta_{C,n}(x) = M_n^H 
\eta_{{\overline C},n}(x)$ and $-i \overline
\sigma^{\mu}\partial_{\mu} \eta_{{\overline C},n}(x) = M_n^H 
\eta_{C,n}(x)$.
This allows us to rewrite Eqs.~(\ref{eq:higgsinoeqs1}) 
-- (\ref{eq:higgsinoeqs4}) as (neglecting singular terms)
\begin{equation}
  M_n^H g_{C,n}(y) - 
    \gamma_{({\overline C},C),n} \partial_{y} g_{C,n}^{c}(y)=0,
\end{equation}
\begin{equation}
  M_n^H g_{{\overline C},n}(y) - 
    \gamma_{(C,{\overline C}),n} \partial_{y} g_{{\overline C},n}^{c}(y)=0,
\end{equation}
\begin{equation}
  M_n^H \gamma_{(C,{\overline C}),n}  
    g_{C,n}^{c}(y)+ \partial_{y} g_{C,n}(y) =0,
\end{equation}
\begin{equation}
  M_n^H \gamma_{({\overline C},C),n} 
    g_{{\overline C},n}^{c}(y) + \partial_{y} g_{{\overline C},n}(y) =0,
\end{equation}
which give solutions
\begin{equation}
  g_{C,{\overline C},n}(y) = N_n \cos(M_n^H y), \hskip 0.5in 
  g_{C,{\overline C},n}^{c}(y) = N_n \sin(M_n^H y),
\end{equation}
with $M_n^H$ defined by
\begin{equation}
  \tan^{2}(M_n^H \pi R) = \frac{\kappa^{2}}{4}.
\end{equation}
Thus, the KK masses for the colored Higgs are given by
\begin{equation}
  M_{n}^H = \frac{1}{R} \left( n + \frac{\arctan(\kappa/2)}{\pi} \right),
\label{eq:higgsspectrum}
\end{equation}
where $n$ runs from negative infinity to positive infinity (the
physical masses are given by the absolute value of $M_n^H$).  For large
$\kappa$, $\arctan(\kappa/2)\simeq \pi/2$, so $M_n^H \simeq (n+1/2)M_C$, 
and the masses fall into nearly degenerate pairs, $(M_n^H, M_{-n-1}^H)$.  
Note that, in contrast to the case of the gauge fields, even for high 
values of $n$, the triplet Higgsinos do not become degenerate with the 
doublet Higgsinos.

Supersymmetry fixes the masses and couplings of the Higgs scalars to
be the same as for the Higgsinos.  The KK towers for the Higgs fields are
shown in Fig.~\ref{fig:spectrum}b.  The doublet spectrum
consists of a pair of massless $N=1$ chiral multiplets, and pairs of
exactly degenerate hypermultiplets at $M_C$, $2M_C$, $3M_C$, and so on.
For large $\kappa$, the triplet spectrum consists of nearly degenerate
pairs of hypermultiplets at $M_C/2$, $3M_C/2$, $5M_C/2$, and so on.
In Fig.~\ref{fig:spectrum}b, the non-degeneracy between these
pairs is not resolved.

\section{Differential running}
\label{sec:diffrunning}

A crucial question in our model is how the gauge couplings evolve
above the compactification scale.  To address this question,
it is useful to focus on the ``differential 
running'', {\it i.e.}, the non-uniform evolution of the gauge couplings. 
Above $M_C/2$ a whole tower of modes contributes to the 
evolution of $g_1,g_2$ and $g_3$. However, we should emphasize that 
{\em this scenario does not employ power law unification.} 
In contrast with Ref.~\cite{Dienes:1998vh}, the overwhelming power-law 
contribution to the running above $M_{C}/2$ is SU(5)
universal, so it is useful 
to focus only on the quantities which distinguish the couplings and lead
to non-uniform evolution above $M_C/2$. For instance, the matter fields fall 
into complete SU(5) multiplets so their effects only change the coupling 
value at unification, but do not influence whether and at what scale the 
couplings unify. In contrast, the presence of the Higgs doublets leads 
to non-uniform running. We will see shortly that above $M_{C}/2$, the 
differential running arises from a sum of a large number of threshold 
effects.\footnote{
The threshold correction from KK towers are also discussed in the 
context of power-law unification \cite{Ghilencea:1998st} and 
orbifold breaking of the unified gauge symmetry \cite{Hall:2001pg}.}

Let us define the scale-dependent quantities
\begin{equation}
  \delta_i(\mu)\equiv \alpha_i^{-1}(\mu)-\alpha_1^{-1}(\mu),
\end{equation}
with $\delta_1$ vanishing trivially.  The evolution of $\delta_i(\mu)$
above $M_C/2$ (but below $m_{\Sigma}$) takes the form
\begin{equation}
  \delta_i(\mu)=\delta_i(M_C/2)-{1\over 2\pi}(R_i^H(\mu)+R_i^G(\mu)),
\end{equation}
where $R_i^H(\mu)$ represents contributions arising from Higgs loops
and $R_i^G(\mu)$ represents contributions arising from gauge
loops. Contributions from matter loops vanish because they contribute
universally to the running.  Using the known spectra for the various
KK towers and their beta function contributions 
(shown in Fig.~\ref{fig:spectrum}), we can calculate
these quantities.  For the Higgs contributions we find 
\begin{equation}
  R_{2}^{H}(\mu) = 
    \frac{2}{5} \log\left({\mu \over M_{C}/2}\right) 
  + \frac{4}{5} \sum_{0<n M_{C}<\mu} \log\left({\mu \over n M_{C}}\right)
  - \frac{2}{5} \sum_{|M^{H}_{n}|<\mu} 
      \log\left({\mu \over |M^{H}_{n}|}\right), 
\end{equation}
\begin{equation}
  R_{3}^{H}(\mu) = 
  - \frac{3}{5} \log\left({\mu \over M_{C}/2}\right) 
  - \frac{6}{5} \sum_{0<n M_{C}<\mu} \log\left({\mu \over n M_{C}}\right)
  + \frac{3}{5} \sum_{|M^{H}_{n}|<\mu} 
      \log\left({\mu \over |M^{H}_{n}|}\right).
\end{equation}
Here the $M_n^H$ are the colored Higgs masses given in 
Eq.~(\ref{eq:higgsspectrum}). Recall that there are actually two slightly
non-degenerate $|M_n^H|$ near each explicitly shown colored Higgs
level in Fig.~\ref{fig:spectrum}b.

For the gauge contributions we have
\begin{equation}
  R_{2}^{G}(\mu) = 
  - 6 \log\left({\mu \over M_{C}/2}\right) 
  - 4 \sum_{0<n M_{C}<\mu} \log\left({\mu \over n M_{C}}\right)
  + 4 \sum_{M^{G}_{n}<\mu} \log\left({\mu \over M^{G}_{n}}\right),
\label{eq:r2g}
\end{equation}
\begin{equation}
  R_{3}^{G}(\mu) = 
  - 9 \log\left({\mu \over M_{C}/2}\right) 
  - 6 \sum_{0<n M_{C}<\mu} \log\left({\mu \over n M_{C}}\right)
  + 6 \sum_{M^{G}_{n}<\mu} \log\left({\mu \over M^{G}_{n}}\right),
\label{eq:r3g}
\end{equation}
where $M_n^G$ are the $X,Y$ vector multiplet masses, given by the
solutions to Eq.~(\ref{eq:gaugespectrum}).

\subsection{Higgs contributions}
The running of $\delta_i (\mu)$ is given by
\begin{equation}
{d\delta_i \over d\mu}=-{1\over2\pi}\left({d R_i^{H}(\mu)\over
  d\mu}+{dR_i^{G}(\mu)\over d\mu}\right).
\end{equation}
Let us consider the Higgs and gauge contributions to this running
separately,
starting with the Higgs loops first.  We have
\begin{equation}
{dR_2^H\over d\mu}
  = {1\over 5\mu}\Bigl[({\rm \#\> of\> doublets\> with\> mass\>
    <\mu})-({\rm \#\> of\> triplets\> with\> mass <\mu}) \Bigr],
\end{equation}
and
\begin{equation}
{dR_3^H\over d\mu}
  = -{3\over 10\mu}\Bigl[({\rm \#\> of\> doublets\> with\> mass\>
    <\mu})-({\rm \#\> of\> triplets\> with\> mass <\mu}) \Bigr].
\end{equation}
Here we mean the number of doublet and triplet chiral superfields:
there are two massless doublets, four triplets with mass $\simeq
M_C/2$, four doublets with mass $M_C$, and so on.

Both of the above expressions have the expected $1/\mu$
characteristic of logarithmic running, but have coefficients which
{\em average out to zero}.  Below $M_C/2$ only the massless $H_u$ and
$H_d$ doublets contribute, so $R_2^H$ ($R_3^H$) runs in the positive
(negative) direction.  At $M_C/2$ we encounter four triplet chiral
superfields and $R_2^H$ ($R_3^H$) now runs in the negative (positive)
direction.
The directions reverse again when we gain four doublets at $M_C$, and
again at $3M_C/2$ when we gain another four triplets.  These threshold
effects continue, but become increasingly negligible.  Taking the
triplet masses to be $(n+1/2 \pm \delta)M_C$, the threshold effects 
from the states between 
$n M_{c}$ and $(n+1)M_{C}$ are proportional to
\begin{eqnarray}
&& \log\left({\Lambda\over n M_{C}}\right) + \log\left({\Lambda
      \over (n+1) M_{C}}\right) - 
  \log\left({\Lambda \over (n+1/2+\delta)M_C}\right)
  -\log\left({\Lambda \over (n+1/2-\delta) M_{C}}\right)
\nonumber\\
&& \qquad\qquad\qquad\qquad\qquad
  =\log\left(\frac{(n+1/2+\delta)(n+1/2-\delta)}{n(n+1)}\right)
\end{eqnarray}
which vanishes as $(1/4-\delta^2)/n^{2}$ for large 
$n$.\footnote{Here we use the step-function approximation for RG 
running. However, had we used a Jacobi $\vartheta$ function as in
Ref.~\cite{Dienes:1998vh}, we would still find the high mass 
threshold effects to be negligible.}

We conclude that 
the differential running due to the Higgs fields dies off quickly
above the compactification scale.  Thus the total Higgs contribution
to the running between $M_C/2$ and a high scale such as $m_{\Sigma}$ or
$g_5^2 \vev{\Sigma}^2$ is essentially independent of the high scale.
Taking the pairs of triplet hypermultiplets to be exactly degenerate,
we find that for high scales,
\begin{eqnarray}
  R_2^H &=& 
  -\frac{2}{5} \left( \log\left({M_C\over M_C/2}\right)-
  \log\left({3M_C/2\over M_C}\right)+\log\left({2M_C\over
      3M_C/2}\right)-\log\left({5M_C/2\over 2M_C}\right)+\dots\right)
\nonumber\\
  &=& -\frac{2}{5} \log\left(\frac{\pi}{2}\right),
\label{eq:higgsrun2}
\end{eqnarray}
and similarly,
\begin{equation}
  R_3^H = \frac{3}{5}\log\left(\frac{\pi}{2}\right).
\label{eq:higgsrun3}
\end{equation}
Note that the ratio $R_2^H/R_3^H=-2/3$ is the same as one would have
calculated by including the contributions from the massless doublets
alone.  This fact will be important when we compare with the running
in 4D minimal SU(5) in section \ref{subsection:uamsu5}.

\subsection{Gauge contributions}
In contrast to the Higgs threshold effects, those arising from 
gauge field loops do not quickly die off. In fact, we will see that 
these effects add up to an effective logarithmic running 
all the way up to $m_{\Sigma}$ or $g_5^2 \vev{\Sigma}^2$.

The definitions we have used for $R_{i}^{G}$ are very convenient 
for making connection to ordinary, four-dimensional running. In 
particular, $\alpha_{1}$ receives no gauge contribution below the 
GUT scale in four-dimensional theories, and likewise here $R_1^G$ will not 
receive contributions to differential running. $R_{2}^G$ and 
$R_{3}^G$, on the other hand, {\em will} receive corrections above the 
compactification scale.

Referring to Eqs.~(\ref{eq:r2g}) and (\ref{eq:r3g}), we see that
the quantities $R_i^G$ are proportional to the quadratic Casimir 
coefficients $C_2(G)$ of the gauge groups.  Thus, 
we can write the gauge contributions as
\begin{equation}
  R_i^G(\mu)=-C_2(G_i)\Delta(\mu),
\label{eq:Delta}
\end{equation}
where $C_2(G)$ is 2 for SU(2) and 3 for SU(3).  Furthermore, we have
\begin{equation}
{d \Delta(\mu)\over d \mu}= {1\over \mu}\Bigl[3+2\Bigl({\rm \#\> of\>
    ({\it V},\Phi)_{3-2-1}\> levels\> below\>\mu }\Bigr)-2\Bigl({\rm \#\> of\>
    ({\it V},\Phi)_{{\it X,Y}}\> levels\> below\>\mu }\Bigl) \Bigr].
\end{equation}
The notation here is as in Fig.~\ref{fig:spectrum}a.  Note that
in contrast to the case for the Higgs multiplets, here the coefficient
multiplying $1/\mu$ does not average to zero.  Rather, for low
modes (such that $M_n^G\simeq (n+1/2)M_C$), the coefficient is 3 up to
$M_C/2$, 1 between $M_C/2$ and $M_C$, then 3 again until $3M_C/2$, and
so on.  For the higher mass modes, such that the $X,Y$ and 3-2-1
multiplets are nearly degenerate, the coefficient becomes fixed at 1.

If the running were coming entirely from the massless 3-2-1 vector
multiplet, the coefficient multiplying $1/\mu$ would be 3, so
we see that above the compactification scale, the differential running
due to the gauge loops is slowed somewhat relative to the ordinary
logarithmic evolution in 4D.  Note that even once the $X,Y$
and 3-2-1 multiplets become degenerate, the differential running continues. 
In this regime, the running is due entirely to the eaten Goldstone states
contained in the $X,Y$ KK tower.  The differential running stops
completely only once we reach the scale $m_{\Sigma}$, when the rest of
the $\Sigma$ degrees of freedom begin to propagate in the
loops.\footnote{Or, if $m_{\Sigma}$ happens to be below the the scale
  at which the $X,Y$ and 3-2-1 towers become degenerate (determined by
  $g_5^2 \vev{\Sigma}^2$), the differential running due to the
  Goldstone degrees of freedom ceases above $m_{\Sigma}$, but the
  total differential running does not die off until a scale $\sim
  g_5^2 \vev{\Sigma}^2$.}

\begin{figure}
\begin{center} 
\begin{picture}(320,200)(-20,-40)
  \Line(0,150)(290,150)
\DashLine(200,136.7)(240,150){3}
\DashLine(200,125)(240,150){3}
\Line(0,70)(200,136.7)
\Line(0,0)(200,125)
\Line(200,136.7)(290,150)
\Line(200,125)(290,150)
\Text(-10,-5)[b]{$\delta_3$}
\Text(-10,65)[b]{$\delta_2$}
\Text(-10,145)[b]{$\delta_1$}
\Line(290,150)(315,150)
\LongArrow(-5,-20)(320,-20)
\Text(320,-35)[b]{$\mu$}
\Line(0,-24)(0,-16)
\Line(200,-24)(200,-16)
\Line(240,-24)(240,-16)
\Line(290,-24)(290,-16)
\Text(0,-38)[b]{$M_Z$}
\Text(200,-38)[b]{$M_C$}
\Text(240,-38)[b]{$M_{\rm GUT}$}
\Text(290,-38)[b]{$g_5^2 \vev{\Sigma}^2$}
  \end{picture}
\caption{The qualitative picture for gauge coupling unification
in our 5D model.  We define
$\delta_i(\mu) \equiv \alpha^{-1}_i(\mu)-\alpha^{-1}_1(\mu)$.
The conventional unification scale $M_{\rm GUT}\sim 2 \times 10^{16}$ GeV
is a derived scale rather than a physical one.  
Here we assume $m_{\Sigma}<g_5^2
\vev{\Sigma}^2$, so that unification is achieved near $g_5^2 \vev{\Sigma}^2$.}
\label{fig:unification}
\end{center}
\end{figure}
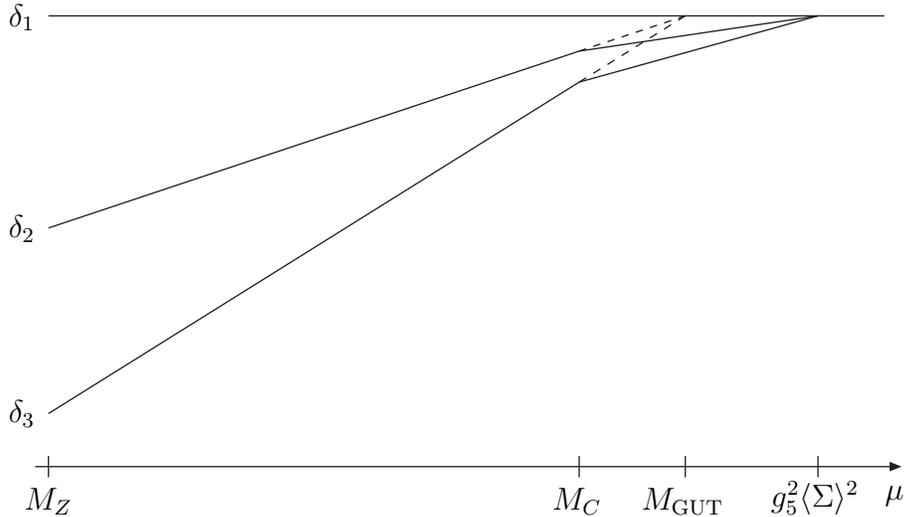

The qualitative picture for gauge coupling unification in our model is
depicted in Fig.~\ref{fig:unification}.  Above the compactification
scale, the non-uniform evolution of the gauge couplings slows, so that
unification occurs at a larger scale than usual.  The unification
scale will essentially be the larger of $m_{\Sigma}$ and
$g_{5}^2\vev{\Sigma}^2$.  Next we will verify that this picture is in
fact correct, and we will make more precise the connection to
unification in ordinary minimal SU(5) in 4D.

\subsection{Unification and minimal SU(5)}
\label{subsection:uamsu5}
Up to this point we have discussed the differential evolution of the 
couplings above the compactification scale, but have not explicitly 
demonstrated that the couplings in our model unify in a manner consistent 
with electroweak scale measurements of the 3-2-1 couplings. Here we 
will demonstrate that, at one loop, the successful prediction of 
$\sin^2 \theta_{W}$ of minimal SU(5) with $M_{H_{C}}<M_{\rm GUT}$ 
\cite{Hisano:1992mh} is equivalent to the 
successful prediction of $\sin^2 \theta_{W}$ in this model with 
$M_{C}\sim  M_{H_{C}}$.  That is, the fact that minimal SU(5) with a 
triplet Higgs below the GUT scale can give $\sin^2 \theta_{W}$ correctly 
{\em guarantees} that this model can do as well.

To see this we must reexamine the running of gauge couplings in 
four-dimensional, minimal SU(5). Above the Higgs triplet mass, the 
gauge couplings evolve as
\begin{equation}
  \alpha^{-1}_{i}(\mu) = 
  \alpha^{-1}_{i}(M_{H_{C}}^{(4d)})-{1\over 2\pi}\left(7-3C_{2}(G_{i})\right) 
  \log \left(\frac{\mu}{M_{H_{C}}^{(4d)}}\right).
\end{equation}

To make connection with our discussion of differential running in the 
five-dimensional theory, we subtract off a universal piece
$\alpha_1^{-1}$ from each gauge coupling, and we find
\begin{equation}
        \delta_{i}^{4d}(M_{\rm GUT})-\delta_{i}^{4d}(M_{H_C}^{(4d)}) =
        {3\over 2\pi}C_{2}(G_{i}) 
        \log\left({M_{\rm GUT}\over M_{H_C}^{(4d)}}\right),
\label{eq:4dunif}
\end{equation}
where $M_{\rm GUT}$ is the conventional GUT scale $\sim 2\times 10^{16}$ GeV.
Below the scale $M_{H_C}^{(4d)}$, the running of the gauge couplings
in the 5D theory is the same as for the 4D theory.  Thus, if we can
demonstrate that $\delta_{i}^{5d}(\mu)-\delta_{i}^{5d}(M_{H_C}^{(4d)})$
has the same form as Eq.~(\ref{eq:4dunif}), we will have shown
that successful unification in ordinary minimal SU(5) implies
successful unification in our model.  Here the superscript in
$M_{H^C}^{(4d)}$ is to emphasize that it is the colored
Higgs mass that gives successful unification in the 4D theory that is
meant, rather than the lightest KK mode of the colored Higgs in the 5D
theory.

The Higgs contributions to
$\delta_{i}^{5d}(\mu)-\delta_{i}^{5d}(M_{H_C}^{(4d)})$ are
\begin{equation}
  -{1 \over 2\pi} c_i \log\left( {M_C/2 \over M_{H_C}^{(4d)}} \right)
  +{1 \over 2\pi} c_i \log\left( \frac{\pi}{2} \right), 
\end{equation}
where $(c_2,c_3)=(2/5,-3/5)$.  The first term comes from the ordinary
running due to $H_u$ and $H_d$ below $M_C/2$, and the second term is
the finite running
above $M_C/2$ found in Eqs.~(\ref{eq:higgsrun2}) and
(\ref{eq:higgsrun3}). Since the two terms are both proportional to
$c_i$, we see that the Higgs contribution to the differential running
in 4D (which takes place entirely below $M_{H_C}^{(4d)}$) is emulated
in 5D with the proper choice of compactification scale.  Had the
differential running due to the Higgs multiplets turned off right at
$M_C/2$, we would simply identify $M_{C}/2=M_{H_C}^{(4d)}$.  Because
we find a small amount of differential running above $M_C/2$, we
instead identify $M_{C}/\pi=M_{H_C}^{(4d)}$. 
We expect small additional corrections to this quantity from further finite 
one-loop effects, but 
the precise value $M_C/\pi$ is not important: 
what matters is that, given
any value of $M_{H_C}^{(4d)}$ that leads 
to a successful prediction of $\sin^2\theta_W$
in the 4D theory, there is a value for $M_C$ ($\sim \pi
M_{H_C}^{(4d)}$) such that the total differential running due to Higgs
loops in the 5D theory will be the same as in the 4D theory.

Having chosen this value for $M_C$,
$\delta_{i}^{5d}(\mu)-\delta_{i}^{5d}(M_{H_C}^{(4d)})$ arises entirely
from loops of gauge and physical $\Sigma$ fields:
\begin{equation}
\delta_{i}^{5d}(\mu)-\delta_{i}^{5d}(M_{H_C}^{(4d)})
  = {3\over2\pi}C_{2}(G_i)\log\left({M_C/2\over M_{H_C}^{(4d)}}\right)
   +{1\over2\pi}C_2(G_i)
   \left\{\Delta(\mu)-\theta(\mu-m_\Sigma)
   \log\left(\frac{\mu}{m_{\Sigma}}\right)\right\},
\end{equation}
where the first term comes from the ordinary 4D running below $M_C/2$, 
and two terms in the curly bracket represent the contributions from 
gauge KK towers and physical $\Sigma$ fields, respectively (for the 
parameterization of the gauge contribution, see Eq.~(\ref{eq:Delta})).
We find that this has the same form as Eq.~(\ref{eq:4dunif}), with the
required identification
\begin{equation}
  \left\{\Delta(\mu)-\log\left(\frac{\mu}{m_{\Sigma}}\right)
  \right\}_{\mu \rightarrow \infty}
  = 3 \log\left({M_{\rm GUT}\over M_C/2}\right).
\label{eq:connect}
\end{equation} 
Note that $\Delta(\mu) \rightarrow \log(\mu/M)$ under $\mu\rightarrow\infty$,
where $M$ is some mass scale which is a function of $M_C$ and 
$g_5^2 \vev{\Sigma}^2$, so that the quantity in the left-hand side has a 
well-defined value which is a function of $M_C, g_5^2 \vev{\Sigma}^2$ 
and $m_{\Sigma}$. ($g_5^2 \vev{\Sigma}^2$ is the scale at which the 
differential running almost ceases in the 5D theory.)

We see that {\em the form of differential running above the Higgs 
triplet mass in minimal SU(5) is precisely what arises from threshold 
effects in the case of the five dimensional theory.} 
This holds irrespective of whether the 
step-function approximation of fields in the RG running is very good 
or not. For instance, had we used the full one-loop calculation as in 
Ref.~\cite{Dienes:1998vh}, all gauge contributions would 
still be proportional to $C_{2}(G)$. 
At one loop, the structure of unification is 
equivalent to that of minimal SU(5). 
That this model can successfully yield $\sin^2 \theta_{W}$ 
is no more nor less remarkable than in four-dimensional unified theories. 
The remarkable feature is that the grand unified group is broken 
``maximally'', at a scale well above the masses of the lightest $X$ and 
$Y$ gauge bosons.  This feature changes the predictions of the theory, 
as we will see in sections \ref{sec:protondecay} and \ref{sec:yukawas}.

Let us make one final comment regarding the differential running of the 
couplings. In a sense, its logarithmic (as opposed to power law) 
behavior is natural. The power law evolution is characteristic of the 
bulk ({\it i.e.}, five dimensional) nature of the gauge coupling. In 
contrast, the SU(5) breaking occurs strictly on a brane. Simply by 
Lorentz invariance, the brane breaking cannot contribute to the bulk 
coupling, but instead should just generate a brane contribution to the 
gauge kinetic term. Consequently, we would not {\em a priori} 
have expected the effects of this SU(5) breaking to 
exhibit a power law behavior. 

\subsection{Scales}
Now that we have shown the direct correspondence between the
differential running in our model 
to that of ordinary minimal SU(5), we can
determine the scale at which successful unification occurs. 
In other words, given a compactification scale $M_{C}$, 
we can find the  values of $m_{\Sigma}$ and $\vev{\Sigma}$ that give 
$\sin^2\theta_W$ correctly.

\begin{figure}
\vspace{-2in}
  \centerline{
    \psfig{file=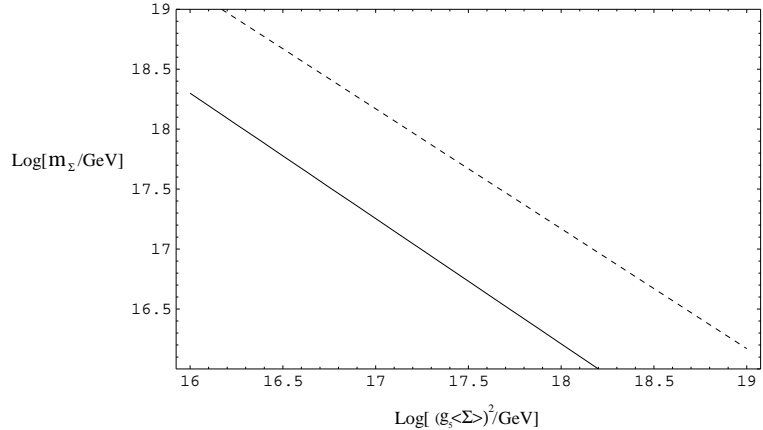,width=.8\textwidth,angle=0} }
\vspace{-2in}
\caption{Values of $M_{\Sigma}$ and $g_{5}^2 \vev{\Sigma}^2$ 
  which generate the same threshold effect above $M_{C}$  
  as is generated in minimal supersymmetric SU(5) between 
  $M_{H_{C}}$ and $M_{\rm GUT}$. 
  The dashed line corresponds to $M_{C}=6\times 10^{14}\gev$, 
  while the solid line corresponds to $M_{C}=6\times 
  10^{15}\gev$. To calculate these lines we have used first-order 
  solutions in 
  $(g_{5}^{2} \vev{\Sigma}^{2} R)^{-1}$ for the $X$ and $Y$ masses.}
\label{fig:allowedscales}
\end{figure}

The allowed compactification scales are already known: these are
related to the allowed values of $M_{H_C}$ in minimal SU(5) by
$M_C\sim\pi M_{H_C}$.  Then we need only to find the values of 
$m_{\Sigma}$ and $\vev{\Sigma}$ for which Eq.~(\ref{eq:connect}) holds. 
These values are indicated in Fig.~\ref{fig:allowedscales} for
compactification scales between $6\times 10^{14}\gev$ 
and $6 \times 10^{15}\gev$ (corresponding to $M_{H_C}\sim 2 \times
10^{14}$ GeV and $2\times 10^{15}$ GeV).  For these compactification
scales, the 5D Planck scale $M_*$ is $(1 \sim 3)\times 10^{17}$ GeV, and
Fig.~\ref{fig:allowedscales} shows that the GUT-breaking scale can be 
at or near this scale and the unification of three gauge couplings can be 
attained below $M_*$ depending on the parameters of the model.\footnote{
If $m_{\Sigma}$ or $g_5^2 \vev{\Sigma}^2$ is above the 5D Planck scale, 
the gauge unification is not completed in a field theoretic regime.
In this case, our field theoretic treatment is not fully trustable 
above $M_*$ and would have some uncertainties coming from the cutoff 
scale physics.}
In the parameter region where the field theoretic unification works, 
the ratio of the cutoff to the compactification scales is $O(10-100)$, 
and no gauge or Yukawa couplings become nonperturbative below 
the cutoff scale.  Thus, our one-loop treatment of the running is 
well justified.  We also find that $\vev{\Sigma}$ must be somewhat 
smaller than $M_*$ in this parameter region.  This is consistent 
with the observation that the theory is more or less strongly coupled 
at the scale $M_*$ to have $O(1)$ Yukawa and gauge couplings in 4D; 
if the theory is strongly coupled at $M_*$, $\vev{\Sigma}$ is 
naturally $\vev{\Sigma} \simeq M_*/(4 \pi)$, since the superpotential 
giving the vev of $\Sigma$ scales like $M_* \Sigma^2 + 4\pi \Sigma^3$.
Even then, however, the $\Sigma$ couples to the other fields in the 
combination of $4\pi \Sigma/M_*$, so that various operators feel 
order one GUT breaking, since $4\pi \vev{\Sigma}/M_* \sim 1$.
Therefore, we treat as if $\Sigma$ has a vev of order $M_*$ in the 
following discussions, although all the arguments also apply in the 
strongly coupled case, $\vev{\Sigma} \simeq M_*/(4 \pi)$.

\subsection{Brane operators}
We finally comment on effects from brane operators like 
$\int d^2\theta (\Sigma/M_*)^m {\cal W}^\alpha {\cal W}_\alpha 
\delta(y-\pi R)$ where $m=1,2,\cdots$ and 
${\cal W}_\alpha$ is the field strength superfield.
Although it is possible that these operators are somehow suppressed 
at the cut-off scale, from effective field theory point of view it is 
generically expected that they are present with order one coefficients.
They give tree-level splitting of three gauge couplings 
for SU(3), SU(2) and U(1), and could affect the previous analysis 
of the gauge coupling unification.  In particular, since we are 
considering $\vev{\Sigma} \approx M_*$, one might think that they 
give $O(1)$ correction to $\sin^2\theta_W$.
However, we can expect that the effect of these operators are 
actually smaller by making the following observations.

As an example, let us first consider the extreme case where all the 
interactions are strongly coupled at the scale $M_*$.  Then, the 
operators involving the field strength superfield scale as 
\begin{equation}
  {\cal L}_5 = 
  \int d^2\theta \left[ \frac{M_*}{24 \pi^3} 
    {\cal W}^\alpha {\cal W}_\alpha + \delta(y-\pi R) 
    \frac{1}{16 \pi^2} \frac{4 \pi \vev{\Sigma}}{M_*}
    {\cal W}^\alpha {\cal W}_\alpha + \cdots \right],
\end{equation} 
where $4 \pi \vev{\Sigma}/M_*$ is an $O(1)$ quantity. (Strictly speaking, 
$4 \pi \vev{\Sigma}/M_*$ must be somewhat smaller than 1 so that 
all higher dimensional operators involving $(4 \pi \vev{\Sigma}/M_*)^m$ 
do not equally contribute and make the theory unpredictable.)
On integrating over $y$, the zero-mode 4D gauge couplings 
$g_0$ are given as
\begin{equation}
  \frac{1}{g_0^2} \sim \frac{M_* R}{12 \pi^2} + \frac{1}{16 \pi^2}.
\end{equation} 
Here, the first and second terms are SU(5)-preserving and 
SU(5)-violating contributions, respectively.
Since we know that $g_0 \sim 1$, we have to take $M_* R \sim 12 \pi^2$ 
in this strongly coupled case.
This shows that the SU(5)-violating contribution coming from brane 
operators is suppressed by a factor of $1/(16 \pi^2)$ in this case.

In fact, the theory is not truly strongly coupled at the scale $M_*$
in the realistic case discussed in previous sections so that the 
one-loop treatment of gauge coupling evolutions is reliable.
Nevertheless, the above argument applies more generically; the 
SU(5)-violating brane contribution is small relative to the 
SU(5)-preserving bulk contribution due to the large volume factor 
$2 \pi R M_*$.  Thus, the correction to $\sin^2\theta_W$ is expected 
to be small.  We will not discuss possible effects of the brane 
operators further, and assume that they are negligible 
in the subsequent discussions.

\section{Proton decay}
\label{sec:protondecay}

One of the key signals of grand unification is proton decay. $X$ and $Y$ 
gauge boson exchange generates dimension six proton decay operators in 
the low energy theory, and Higgsino triplet exchange generates dimension 
five operators. One might expect that since the particles appear at a 
scale $\sim M_{C}$, this will be the suppression scale of the proton decay 
operators. One interesting possibility offered by this framework is that 
{\em the rate of proton decay is not constrained by gauge invariance to 
be related to the strength of gauge and Yukawa interactions in the
usual way.}

We have already noted that the dangerous particles have 
wave functions at the $y=\pi R$ fixed point
that are small compared with those of their 3-2-1 or doublet 
counterparts. On the other hand, there is a tower of states that 
can mediate proton decay, and we must sum each contribution.
In this section, we investigate proton decay operators generated by 
exchanges of $X, Y$ gauge bosons and colored Higgsinos, and show that 
the present model can satisfy the constraints coming from experimental 
lower bounds on the proton lifetime.

\subsection{Dimension six operators}
Dimension six proton decay operators arise from the exchange of 
$X$ and $Y$ gauge bosons. The coupling of these bosons to fields on the 
brane is suppressed by a factor $\cos(M_n^G \pi R)$. Comparison with 
four-dimensional theories can be made by replacing
\begin{equation}
  \frac{1}{M_{X,Y}^{2}}\Longrightarrow 2 \sum_{n=0}^\infty 
    \frac{\cos^{2}(M_{n}^G \pi R)}{{M_{n}^G}^{2}},
\end{equation}
where $M_n^G$ are the masses of the $X$ and $Y$ gauge boson KK modes.
These masses satisfy Eq.~(\ref{eq:gaugespectrum}), which allows us
to approximate the sum as
\begin{equation}
  \sum_{n=0}^\infty \frac{4}{g_{5}^{4} \vev{\Sigma}^4+ M_{C}^{2} 
    (2n+1)^{2}} = \frac{\pi \tanh(\pi g_{5}^{2} \vev{\Sigma}^{2}/2 
    M_{C})}{g_{5}^{2}\vev{\Sigma}^{2} M_{C}}.
\end{equation}
Thus, the ``effective mass'' of the $X$ and $Y$ bosons is 
$\approx g_{5}\vev{\Sigma} M_{C}^{1/2}/(2\pi)^{1/2}=g_4
\vev{\Sigma}$. For  $\vev{\Sigma}$ at or near the five dimensional 
Planck scale, this will typically be larger than $M_{\rm GUT}$, 
although it is not required to satisfy experimental constraints.
In any case, despite the fact that the lightest $X$ and $Y$ gauge
bosons have masses of only $\sim 10^{15}$ GeV, it is easy to
satisfy proton decay constraints
coming from dimension six operators in this model.

\subsection{Dimension five operators}
Dimension five operators come from integrating out the Higgsino 
triplets. Again, we make connection with the four dimensional theory 
by finding an effective triplet Higgsino mass by taking the whole sum
over KK modes. Here we make the replacement
\begin{equation}
  \frac{1}{M_{H_{C}}} \Longrightarrow \sum_{n=-\infty}^{\infty} 
    \frac{\cos^{2}(M_{n}^H \pi R)}{M_{n}^H},
\end{equation}
where $M_n^H$ are the masses of the colored Higgs KK modes, given in
Eq.~(\ref{eq:higgsspectrum}). Using this spectrum, we obtain the sum
\begin{equation}
  \sum_{n=-\infty}^{\infty} \left(\frac{4}{\kappa^{2}+4}\right)
    \frac{1}{n M_{C}+ M_{C} \arctan(\kappa/2)/\pi} 
  = \frac{8 \pi}{\kappa (\kappa^{2}+4) M_{C}}
\end{equation}
Thus, proton decay from dimension five operators is suppressed by a 
factor $\sim 8 \pi/\kappa^{3}$ compared to the compactification 
scale.  In ordinary minimal SU(5), proton decay limits require
$M_{H_C}>9.4 \times 10^{16}$ GeV, so for a compactification scale of
$M_{C}=2 \times 10^{15}\gev$, we would need 
a $\kappa$ parameter of ten to adequately suppress proton decay 
operators. This leaves open the possibility that proton decay could 
be observed in future experiments.
However, for larger $\kappa$ detection becomes increasingly unlikely.

\subsection{Derivative operators}
In addition to the ordinary Yukawa couplings between the Higgs and 
matter fields on the brane, there can be derivative couplings of 
the conjugate Higgs fields to the matter fields as well.\footnote{
We thank M. Graesser for bringing this to our attention.}
These operators such as $(\zeta_u T_{\bf 10} T_{\bf 10}\, \partial_{y} 
H_{\overline{\bf 5}}^{c} + \zeta_d T_{\bf 10} F_{\overline{\bf 5}} 
\, \partial_{y} H_{\bf 5}^{c}) \delta(y-\pi R)$ can lead to proton decay.
Of course, coefficients $\zeta_u$ and $\zeta_d$ of these operators are 
not related to the usual Yukawa couplings, $\lambda_u$ and $\lambda_d$, 
by SU(5), so that their actual significance is unknown. 
Furthermore, there is an ambiguity in what we mean by the derivative 
of the conjugate field, which is not differentiable at the 
point $y=\pi R$. However, using Eqs.~(\ref{eq:higgsinoeqs1}) and 
(\ref{eq:higgsinoeqs2}), we can rewrite the coupling as
\begin{equation}
  \delta(y-\pi R) \partial_{y} H_{C}^{c} 
  = \delta(y-\pi R) \sum_{n} \left( M_{n}^{H} 
    g_{C,n}(\pi R) - \frac{\kappa}{2 \pi R} \sum_{m} 
    g_{\overline C,n}(\pi R) \right) \eta_{\overline C,n}(x),
\end{equation}
where the summation over $m$ arises from rewriting the 
$\delta$-function as a sum of the KK mode.

While we can compute proton decay diagrams involving these vertices, 
it is now apparent that such diagrams have a strong dependence on 
how we cut off the sum of the KK mode.  If the cutoff is done near 
the fundamental scale, these diagrams can, at least in principle, 
give comparable contribution to those involving the usual Yukawa 
couplings. However, it is not entirely clear what the 
cutoff for these diagrams should be. In the case where the 
brane is dynamical, the summation in the above operators are cut off 
at the scale of the brane tension \cite{Bando:1999di}.
For couplings at the orbifold fixed point, which is not dynamical and 
cannot fluctuate, it is unclear whether the cutoff is the fundamental 
scale or a lower scale, such as the radion mass.

For our purposes here, we will not address these issues further. Since 
we cannot {\em a priori} know the size of the couplings of these 
operators and their flavor structure, estimating the resulting proton 
decay rate is already very uncertain. However, it is possible that 
such operators may provide an opportunity for detectable dimension 
five proton decay in the near future.

\section{Yukawa couplings}
\label{sec:yukawas}

So far, we have a framework that looks like ordinary SU(5) except with
suppressed proton decay. While the breaking occurs at a scale near the
5D Planck scale $M_*$, the $X, Y$ gauge bosons are much lighter. 
Still, in a very real sense, the breaking of SU(5) is ``maximal'', 
which can manifest itself in deviations from SU(5) expectations. 
We here consider the effects of such maximal GUT breaking on the 
fermion Yukawa couplings.

In the minimal SU(5), one important prediction is the unification 
of the Yukawa couplings.  A successful prediction of the theory 
is the unification of the bottom and $\tau$ Yukawa couplings 
at the GUT scale \cite{Chanowitz:1977ye}.  However, it is well known 
that the SU(5) relations fail in the lighter first-two generations.
For instance, an SU(5) relation $m_{e}/m_{\mu}=m_{d}/m_{s}$ fails 
by a factor of ten.\footnote{One may be able to correct this prediction 
by means of contributions from supersymmetry-breaking $A$ terms 
to the Yukawa couplings \cite{Diaz-Cruz:2000mn}.}

In ordinary 4D SU(5) GUT, it has been suggested that the operators 
involving the $\Sigma$ field can correct this discrepancy 
\cite{Ellis:1979fg}.  However, this mechanism does not work in most 
theories where the fermion mass hierarchy is explained by the
Froggatt-Nielsen mechanism \cite{Froggatt:1979nt}.  In this mechanism, 
the matter fields carry generation dependent flavor U(1) charges and 
the various Yukawa couplings are generated through the U(1) breaking 
spurion.  This is an attractive mechanism in that it not only suppresses 
the first-two generation Yukawa couplings but also suppresses dangerous 
tree-level dimension five proton decay operators.  However, having
employed this mechanism, the GUT-breaking operators involving 
$\vev{\Sigma}$ can modify SU(5) mass relations only by an 
amount suppressed by a factors of $M_{\rm GUT}/M_{\rm Pl}$, which 
is too small to accommodate order one deviations suggested by the 
observed quark and lepton masses.\footnote{One possible way to evade 
this conclusion is to assign a non-vanishing Froggatt-Nielsen charge 
to the $\Sigma$ field \cite{Izawa:1999cm}.}

In the present model, on the other hand, there is no suppression of 
higher dimensional operators involving GUT-breaking effects, since 
$\vev{\Sigma}$ is near the cut-off scale $M_*$.  In this sense, 
fields living on the GUT-breaking brane see the GUT broken maximally, 
and their Yukawa couplings need not respect the SU(5) symmetry. 
Thus, the failure of SU(5) to describe $(m_{e}/m_{\mu})/(m_{d}/m_{s})$ 
seems quite natural.  In this framework, however, the success of the 
$\lambda_{b}$-$\lambda_{\tau}$ unification must be viewed as an 
accident, unless there is a reason why the coupling of $\Sigma$ 
to the third generation is somewhat suppressed.

\section{Larger gauge groups}
\label{sec:SO(10)}

Up to this point, we have restricted our attention to a scenario with 
SU(5) gauge group, but it is interesting to consider larger gauge groups. 
One possible difference comes from additional matter fields required 
by larger gauge groups.  For instance, in SO(10) there is an additional 
state, right-handed neutrino.  If this additional field acquires a mass 
around the cut-off scale $M_{*}$ from SO(10) breaking, then there will 
be a running effect between $M_{C}$ and $M_{*}$ which is not SO(10) 
universal.  

It is, however, important that this does not affect our 
previous analyses; the 3-2-1 unification still works as in section 
\ref{sec:diffrunning} even in the case of SO(10).
The additional Higgs states do not contribute to differential 
running above $M_{C}$, and the threshold effects are proportional to 
3-2-1 $\beta$-functions. This means that we can have larger gauge 
groups broken all the way to 3-2-1 around the cut-off scale.
In the SO(10) case, this may result in too large right-handed neutrino 
masses to give an appropriate mass scale for atmospheric oscillations 
\cite{Fukuda:1998mi} through see-saw mechanism \cite{Seesaw}.  
Then, we may need somewhat small coefficient in front of the 
operator which gives right-handed neutrino masses or have to resort 
to other ways to generate neutrino masses within supersymmetric 
models, for instance, though $R$-parity violation \cite{Hall:1984id} 
or supersymmetry breaking \cite{Arkani-Hamed:2000bq}.

\section{Conclusions}

The possibility that the standard model gauge group is merely a 
subgroup of a larger, simple group is an attractive one. 
Unfortunately, the simplest version of supersymmetric SU(5) predicts
proton decay at rates incompatible with experiment. 

We have demonstrated that the incorporation of just one new 
ingredient --- an additional dimension in which gauge and
Higgs fields propagate --- 
brings about crucial changes relative to ordinary GUTs. 
The couplings of the lightest $X$ and $Y$ bosons 
are no longer directly linked 
through gauge symmetry to those of the lightest 3-2-1 bosons, and the 
couplings of the 
lightest Higgs triplets are no longer directly linked to those of 
the Higgs doublets. The strong 
breaking of the GUT symmetry ``pushes away'' the wave functions of 
these states, suppressing the generated proton decay below 
experimental limits, 
all while retaining {\em ordinary, logarithmic unification}. 
Such features seem special to a five dimensional theory, 
incapable of reproduction in simple four dimensional theories.

The presence of extra dimensions of this size has been motivated 
previously as a means to resolve the supersymmetric flavor 
problem \cite{Randall:1999uk}.  Our GUT-breaking picture can nicely 
fit into this framework of supersymmetry-breaking mediation.
For instance, if supersymmetry is broken at $y=0$ fixed point by 
$F$-term vev of singlet or non-singlet field, it naturally realizes 
the scenarios of Refs.~\cite{Kaplan:2000ac} and 
\cite{Kaplan:2000jz}, respectively.  The incorporation of our 
GUT picture within these scenarios may give interesting signatures, 
since we now have one more piece of 
information about parameters of the model; 
the compactification radius is determined by the gauge coupling 
unification.  We leave an investigation of detailed phenomenology 
for future work.

To summarize, the framework we have described is very simple, but 
can describe various observed features such as gauge coupling 
unification, lack of unification in Yukawa couplings, the absence of 
proton decay, and so on.  It will be interesting to add non-minimal 
structure to the model as a means of deriving experimental signatures.

\section*{Acknowledgement}

We thank M.L.~Graesser, L.J.~Hall, D.B.~Kaplan, D.E.~Kaplan, 
A.E.~Nelson, M.E.~Peskin and L.~Randall for useful discussions.

\end{document}